\documentclass[sigconf]{acmart}

\usepackage{multirow}
\usepackage{amsmath}
\usepackage{listings}
\usepackage{enumitem}
\usepackage{xspace}

\setlist[itemize]{leftmargin=0.5cm}
\setlist[enumerate]{leftmargin=0.5cm}

\newcommand{\axd}{arXivDigest\xspace}

\copyrightyear{2020} 
\acmYear{2020} 
\setcopyright{acmcopyright}\acmConference[CIKM '20]{Proceedings of the 29th ACM International Conference on Information and Knowledge Management}{October 19--23, 2020}{Virtual Event, Ireland}
\acmBooktitle{Proceedings of the 29th ACM International Conference on Information and Knowledge Management (CIKM '20), October 19--23, 2020, Virtual Event, Ireland}
\acmPrice{15.00}
\acmDOI{10.1145/3340531.3417417}
\acmISBN{978-1-4503-6859-9/20/10}

\begin{document}

\fancyhead{}
\title{ArXivDigest: A Living Lab for Personalized Scientific Literature Recommendation} 

\author{Kristian Gingstad}
\affiliation{University of Stavanger}
\email{k.gingstad@stud.uis.no}

\author{{\O}yvind Jekteberg}
\affiliation{University of Stavanger}
\email{o.jekteberg@stud.uis.no}

\author{Krisztian Balog}
\affiliation{University of Stavanger}
\email{krisztian.balog@uis.no}


\begin{abstract}
Providing personalized recommendations that are also accompanied by explanations as to why an item is recommended is a research area of growing importance. At the same time, progress is limited by the availability of open evaluation resources. In this work, we address the task of scientific literature recommendation. We present \axd, which is an online service providing personalized arXiv recommendations to end users and operates as a living lab for researchers wishing to work on explainable scientific literature recommendations.
\end{abstract}


\maketitle

\section{Introduction}
\label{sec:intro}

Recent years have seen an increased interest in recommender systems.
Evaluation is a central aspect of research in this area, where the need for both \emph{offline} and \emph{online} evaluation, as complementary approaches, has been recognized~\citep{Beel:2013:CAO,Garcin:2014:OOE}.
Online evaluation, however, is challenging as it requires a live service with sufficient traffic volume, which is generally unavailable to those outside research labs of major service providers.
\emph{Living labs} was proposed as an alternative, where third-party researchers are allowed to replace components of a live service, under certain restrictions, and have real users of the service interact with the generated results~\citep{Hopfgartner:2019:CEL}.
In this paper, we propose a living lab for scientific literature recommendation.

Academic search, as a use case, is appealing for many reasons.  Generally, data is openly available, and there is already a number of services consolidating scientific literature and associated metadata.  Here, we specifically focus on providing recommendations over papers published on arXiv,\footnote{\url{https://arxiv.org/}} which has become a leading outlet for bleeding edge research (especially for machine learning-related work).
Given the accelerating pace at which scientific knowledge is being produced and consolidated on arXiv, it has become a real need to provide a recommendation service that helps researchers to keep up with the articles published there.
Academic search is also interesting from a research perspective, as it provides a fertile ground for current research problems, including, e.g., semantic matching to overcome vocabulary mismatches~\citep{Jagerman:2018:OLL}. 

We acknowledge the multitude of related efforts in this space (cf. Sect.~\ref{sec:related}).  What makes \axd unique that it aims to provide an open service that we, researchers, would enjoy using (thereby subscribing to the ``eat your own dog food'' principle).
It is meant to be an ongoing effort that is shaped and developed in a way that it best serves the community's interests.  One specific example of this is explainability.  Explainable AI has been identified as an increasingly important area of research~\citep{Monroe:2018:AEY,Zhang:2020:ERS}.  However, evaluation of explainable approaches represents a major bottleneck.  Experimentation with live users in commercial services is severely limited due to scalability, quality, and ethical concerns.  As such, they tend to take a conservative stance.
Conversely, most researchers appear to be open regarding their work and research interests, which removes the barriers and issues regarding privacy.  This makes it possible for us to complement recommendations with explanations that users can comment on.
Also, researchers can be both users and developers in \axd, and can thereby enjoy full transparency.

In particular, users subscribing to the \axd service receive personalized article recommendations, which are emailed to them in daily/weekly digests and can also be viewed on a web interface.  Users can leave feedback on the recommendations they receive as well as on the accompanying explanations.  They can further save favourite articles.  All these interactions are registered and used to help generate better recommendations for them in the future.
Researchers can register their own recommender system, by requesting an API key, and get access to profile and interaction data of users.  They can then generate personalized recommendations for users and upload these via the API.  Users will then be exposed to recommendations generated by multiple systems.

The service is available at \url{https://arxivdigest.org/}. The source code and API documentation are published at \url{https://github.com/iai-group/arXivDigest}.
\begin{figure*}[!ht]
    \centering
    \vspace*{-0.5\baselineskip}
    \includegraphics[width=0.8\textwidth]{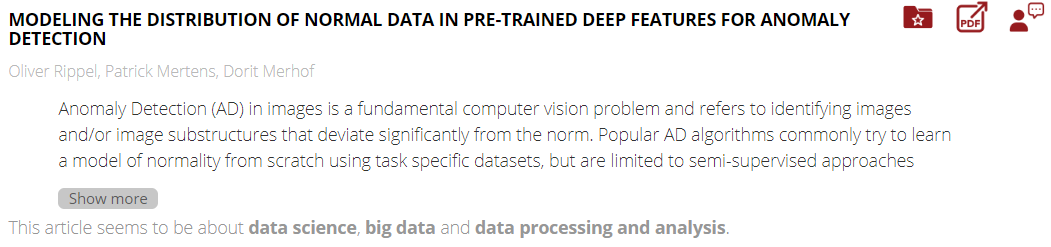}
    \vspace*{-0.5\baselineskip}
    \caption{Article recommendation shown on the web interface.}
    \label{fig:recommendation}
    \vspace*{-0.5\baselineskip}
\end{figure*}

\section{Related Work} 
\label{sec:related}

There are numerous services in the space of academic search, including digital library search engines, such as 
CiteSeerX~\citep{wu-citeseerx-2014} or SSOAR.\footnote{\url{https://www.gesis.org/ssoar/home/}}
There also exist services that consolidate scientific literature and associated metadata, offer API access to these, as well as provide a range of search and recommendation services themselves.
Prominent examples include AMiner,\footnote{\url{https://www.aminer.cn/}} Microsoft Academic Search,\footnote{\url{https://academic.microsoft.com/}} and Semantic Scholar.\footnote{\url{https://www.semanticscholar.org/}}
ArXiv-sanity\footnote{\url{http://www.arxiv-sanity.com/}} is a service specifically for arXiv, helping users to find related articles. 

Benchmarking efforts using living labs include CLEF NewsREEL~\cite{HopfgartnerBSKL15}, which provided an live evaluation platform for the task of news recommendations.
The CLEF LL4IR track~\citep{Schuth:2015:OLL} featured product search and web search as use cases.
The OpenSearch track at TREC~\citep{Jagerman:2018:OLL} addressed the task of ad hoc scientific document retrieval using CiteSeerX and SSOAR as live platforms.
None of these benchmarks offered the possibility for personalization nor for providing explanations.

\section{The Recommender Service}
\label{sec:service}

ArXivDigest is a scientific literature recommendation service that provides users with personalized suggestions based on their interest profile.
By using the service, users agree to `donate' the data they generate for research purposes.  Specifically, their profile information (name, websites, and topics of interest), the recommendations they received, and their interactions with those recommendations are made available to experimental systems via an API (cf. Sect.~\ref{sec:platform:api}).
Users can download all data stored about them from the website, and can remove themselves entirely from the system, as per GDPR.

Below, we provide a brief overview of user-facing functionality.

\begin{figure}[!ht]
    \centering
    \includegraphics[width=0.88\columnwidth]{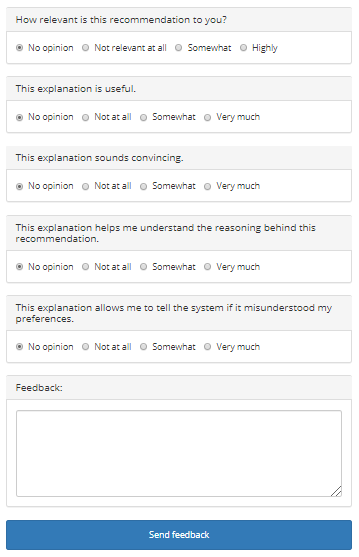}
    \vspace*{-0.75\baselineskip}
    \caption{User feedback form for article recommendations.}
    \label{fig:feedback}
    \vspace*{-0.75\baselineskip}
\end{figure}

\begin{itemize}
	\item \textbf{Sign-up/profile}: In order to make personalized recommendations, we need to have user profiles with personal information.  Therefore, users need to register by filling out a sign-up form where they provide basic details (name and email address), link to their DBLP and/or Google Scholar profile, specify keywords of interest, and choose the regularity of digest emails (daily or weekly).  Users can modify their profile any time later, view all data associated with them, and remove themselves from the system.

	\item \textbf{Article recommendations}: Registered users can view the articles that are recommended to them, either in the digest emails or on the web interface.  See Fig.~\ref{fig:recommendation} for an example.  All recommendations are accompanied by an explanation.  Articles can be saved to a personal library (``liked'') to improve recommendations and for easy future re-finding.  

	\item \textbf{Topic recommendations}: A natural way of representing users' interests is via a set of topics (short natural language phrases).  We aid users in populating their profiles with additional topics of interest, by displaying a list of topic recommendations on the website.  They can accept or reject items in the list with a single click. 

	\item \textbf{Feedback}: Users can leave feedback on the recommendations and/or on the accompanying explanations. For article recommendations, a detailed form is given, asking users about the \emph{relevance} of the recommendation, as well as about how \emph{satisfactory}, \emph{persuasive}, \emph{transparent}, and \emph{scrutabile} they found the explanation (the choice of particular explanation dimensions was informed by~\citep{Balog:2020:MRE}); see Fig.~\ref{fig:feedback}.  Feedback on other aspects of the system (bug reporting and feature requests) is free-text.
\end{itemize}

\section{The Living Lab Platform}
\label{sec:platform}

ArXivDigest operates as a living lab platform, by providing a broker infrastructure that connects researchers that have signed up for the service (\emph{users} for short) and experimental systems that provide content recommendations (\emph{systems} for short).  Systems generate personalized recommendations for all users and make these available to the broker (by uploading them via an API).  The broker takes all recommendations created for a given user, interleaves them, and makes the top-$k$ recommendations available to users.  Further, the broker registers user feedback (and makes it available to systems).  This process is repeated daily.
Specifically, there are two types of items that can be recommended to users: articles (i.e., arXiv papers) and topics (i.e., keywords of interest).
Articles are sent out in a digest email and can also be viewed on the web interface.  Topic recommendations are only available via the web interface.

\subsection{Evaluation Methodology}

We adhere to an \emph{online} evaluation methodology for information retrieval~\citep{Hofmann:2016:OEI}.
Users are presented with a ranked list of (article or topic) recommendations, which is a result of interleaving rankings of multiple systems.  
Specifically, we employ multileaving, which is designed to effectively compare more than two rankers at the same time~\citep{Schuth:2016:SEL}.
By \emph{impression} we mean a combined ranking that is seen by a user (i.e., it counts even if there is no interaction).
There may be zero to multiple \emph{user interactions} associated with each impression.

The following user interactions are distinguished for article recommendations, with associated reward points in parentheses: \emph{saved} to personal library (5), \emph{clicked} in email (3) or on the web (3), and \emph{seen} in email (0) or on the web (0).
For topic recommendations, user interactions (and rewards) are: \emph{accepted} (1), \emph{rejected} (0), \emph{refreshed} (0), and \emph{expired} (0). The last two actions mean that the user has seen the list of recommendations, but did not interact with them. 

In the traditional interleaving setting, where an experimental system is compared against a production system,  the performance of each system is measured in terms of wins/losses based on the clicked results~\citep{Jagerman:2018:OLL}.
In our setting, interactions are not limited to clicks and there are more than two systems that are being compared.  Thus, we introduce a new evaluation metric based on the notion of \emph{Reward}.
The \emph{Reward} of a system $s$ in an interleaving $I$ is defined as the weighted sum of user interactions with results originating from that system.  For example, if a system in an interleaving has received 3 clicks on recommended articles, 2 of which also got saved by user, the reward of this system would be $3 \times 3 + 2 \times 5=19$.
To ensure the comparability of systems, we define \emph{Normalized Reward} as the reward of a system divided by the total reward resulting from that impression.  That is, the normalized rewards of all systems partaking in the interleaving sum up to 1.
Finally, \emph{Mean Normalized Reward} for a system over a set time period is calculated by taking the mean of the Normalized Reward accumulated over the given period.

System performance is monitored continuously over time, with performance indicators (\#impressions and Mean Normalized Reward) made available to system owners via an interactive administration interface.  For comparing a set of systems, performance is to be measured during a designated (and sufficiently long) evaluation period.	
To ensure a fair comparison across systems, our multileaver will select systems at random for each multileaving, but systems that have fewer impressions will be preferred.  This way, all systems can receive approximately the same amount of impressions.

\subsection{Architecture}

\begin{figure}[!t]
    \centering
    \vspace*{-0.25\baselineskip}
    \includegraphics[width=0.8\columnwidth]{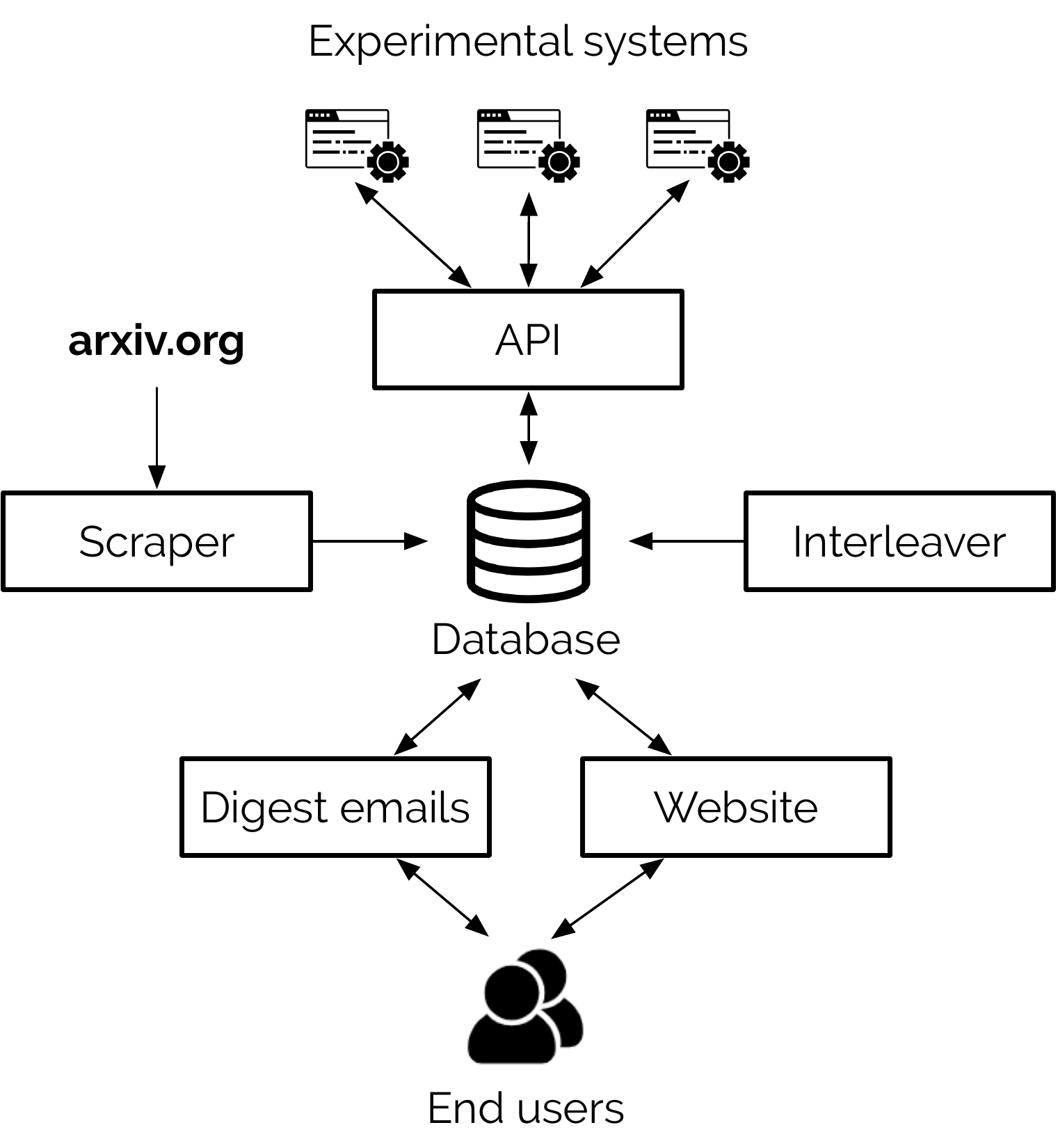}
    \vspace*{-0.75\baselineskip}
    \caption{Architecture of the \axd platform.}
    \label{fig:architecture}
\end{figure}

The main architectural components, shown in Fig.~\ref{fig:architecture}, are an \emph{API} connecting \emph{experimental systems} with the \axd service (detailed in Sect.~\ref{sec:platform:api}), a \emph{scraper} to fetch new articles from arXiv, an \emph{interleaver} to combine results of experimental systems to recommendation lists shown to \emph{end users} either in \emph{digest emails} or on the \emph{web front-end}, and a \emph{database} back-end (MySQL).

All code (except single launching scripts) is contained in a single Python package (\texttt{arxivdigest}), which makes code sharing between the different components easy. Also, installing and updating can be handled by a standard setup script.
The package contains four modules: \texttt{frontend}, \texttt{api}, \texttt{connector} (to facilitate clean and easy communication with the API, and to help reduce the amount of code to be written for each recommender system), and \texttt{core} (code for interleaving, scraping, and email services).  
The web front-end and API are built using Flask\footnote{\url{https://flask.palletsprojects.com/en/1.1.x/}} and are deployed as WSGI applications.
The scraper, interleaver, and digest emails are run as batch processes.

\subsection{The arXivDigest API}
\label{sec:platform:api}

We provide a RESTful API for experimental recommender systems to access article and user data, and to upload personalized article/ topic recommendations to be evaluated with live users.
Developers of said systems first need to request an API key.  To complete the API registration process, they further need to sign the API Terms of Usage, which forbids storing user-specific data for more than 24 hours.  At the same time, data obtained from the API may be displayed or published in a technical or scientific context, provided that specific individuals cannot be identified.

\subsection{Process for Experimental Systems}

Systems are given a 2.5 hour window each day to download new content once it has been published on arXiv and generate recommendations for all registered \axd users.
The specific steps of submitting article recommendations are listed below (topic recommendations follows analogously, but is omitted here in the interest of space).

\begin{enumerate}[label=\arabic*.]
    \item Call \texttt{GET /} to get the settings of the API. 
    \item Call \texttt{GET /users?from=0} to get a batch of user IDs; the offset may be incremented to get new batches. 
	\item Call \texttt{GET /user\_info?ids=...} with the user IDs as a  query parameter, to get information about the users.  Optionally, additional data based on the available user profiles may be gathered from external services.
	\item Call \texttt{GET /articles} to get the IDs of articles that are candidates for recommendation.  These are articles that have been published on arXiv within the last 7 days, to have a sufficiently large pool of articles to recommend from.
	\item Call \texttt{GET /article\_data?article\_id=...} with the article as a query parameter, to get information about a given article.  Optionally, additional metadata may be gathered from external sources (e.g., from Semantic Scholar).
	\item Call \texttt{GET /user\_feedback/articles?user\_id=...} with the user IDs as a query parameter to get information about what recommendations have already been shown to a user. These articles should be filtered out as they will be ignored by the platform.  
	\item Use the available data about users and articles to create personalized recommendations with explanations for each user. Important parts of the explanations may be boldfaced by using markdown-style markup (like \texttt{**text**}).
	\item Call \texttt{POST /recommendations/articles} to submit the generated article recommendations in batches of the size defined by the API settings.
	\item Repeat steps 2 to 8 until all user batches have been given recommendations. 
\end{enumerate}
The above steps are meant to be repeated every weekday, e.g., by setting up some batch process.  This, however, is not enforced.  Systems not submitting recommendations for certain days or users (e.g., if no suitable matches are found) are not penalized in any way other than receiving less `exposure.'  It is worth mentioning that recommendations made for each user are pushed to a stack, and each day the highest scoring ones are taken by the interleaver process. This way, systems have the possibility to update their recommendations.

\section{Baseline Article Recommender}
\label{sec:baseline}

A simple baseline recommender method has been implemented on top of Elasticsearch, and is shipped with the \axd codebase.
For a given user, it scores all candidate articles against each of the user's topics, using a standard retrieval method (BM25).  Then, each article receives the sum of all retrieval scores of all user topics as its final score.
The top-$k$ highest scoring articles are selected as recommendations.
The corresponding explanations are generated by selecting the top-3 highest scoring topics for each article and instantiating the template ``This article seems to be about [$t_1$], [$t_2$] and [$t_3$],'' where [$t_1$], [$t_2$], and [$t_3$] are placeholders for topic names (and are rendered boldfaced, cf. Fig.~\ref{fig:recommendation}).

\section{Conclusion and Future Directions}

We have presented the \axd service and platform for personalized scientific literature recommendation.
At the time of writing, the service is operational and already has a small user base. 
The living lab platform is also up and running for researchers to deploy their own recommendation methods.
In addition to the baseline article recommendation system presented here, a number of more advanced article and topic recommendation approaches have been developed and deployed by the authors of this paper, serving end users with a diverse set of suggestions.  (These experimental systems are not discussed here, as these  are not part of the core platform, but they are linked from the \axd GitHub repository.)

We see the CIKM conference as an major opportunity to talk about our initiative and to get other researchers involved in this project, both as contributors to the \axd platform and API, as researchers developing novel explainable recommender approaches, and as end users using the service.

It is our hope to organize a dedicated track to scientific literature recommendation, using \axd as the living labs platform, in the near future at an international benchmarking campaign (possibly, as a continuation of the TREC Open Search track~\citep{Jagerman:2018:OLL}).
We also see this platform contributing to other related efforts planned within the community, and in particular to the idea of a Scholarly Conversational Assistant, which has been proposed in~\citep{Balog:2020:CCC}.

\bibliographystyle{ACM-Reference-Format}
\bibliography{cikm2020-arxivdigest}


\begin{thebibliography}{13}


\ifx \showCODEN    \undefined \def \showCODEN     #1{\unskip}     \fi
\ifx \showDOI      \undefined \def \showDOI       #1{#1}\fi
\ifx \showISBNx    \undefined \def \showISBNx     #1{\unskip}     \fi
\ifx \showISBNxiii \undefined \def \showISBNxiii  #1{\unskip}     \fi
\ifx \showISSN     \undefined \def \showISSN      #1{\unskip}     \fi
\ifx \showLCCN     \undefined \def \showLCCN      #1{\unskip}     \fi
\ifx \shownote     \undefined \def \shownote      #1{#1}          \fi
\ifx \showarticletitle \undefined \def \showarticletitle #1{#1}   \fi
\ifx \showURL      \undefined \def \showURL       {\relax}        \fi
\providecommand\bibfield[2]{#2}
\providecommand\bibinfo[2]{#2}
\providecommand\natexlab[1]{#1}
\providecommand\showeprint[2][]{arXiv:#2}

\bibitem[\protect\citeauthoryear{Balog, Flekova, Hagen, Jones, Potthast,
  Radlinski, Sanderson, Vakulenko, and Zamani}{Balog et~al\mbox{.}}{2020}]%
        {Balog:2020:CCC}
\bibfield{author}{\bibinfo{person}{Krisztian Balog}, \bibinfo{person}{Lucie
  Flekova}, \bibinfo{person}{Matthias Hagen}, \bibinfo{person}{Rosie Jones},
  \bibinfo{person}{Martin Potthast}, \bibinfo{person}{Filip Radlinski},
  \bibinfo{person}{Mark Sanderson}, \bibinfo{person}{Svitlana Vakulenko}, {and}
  \bibinfo{person}{Hamed Zamani}.} \bibinfo{year}{2020}\natexlab{}.
\newblock \showarticletitle{Common Conversational Community Prototype:
  Scholarly Conversational Assistant}.
\newblock \bibinfo{journal}{\emph{CoRR}}  \bibinfo{volume}{abs/2001.06910}
  (\bibinfo{year}{2020}).
\newblock


\bibitem[\protect\citeauthoryear{Balog and Radlinski}{Balog and
  Radlinski}{2020}]%
        {Balog:2020:MRE}
\bibfield{author}{\bibinfo{person}{Krisztian Balog} {and}
  \bibinfo{person}{Filip Radlinski}.} \bibinfo{year}{2020}\natexlab{}.
\newblock \showarticletitle{Measuring Recommendation Explanation Quality: The
  Conflicting Goals of Explanations}. In \bibinfo{booktitle}{\emph{Proc. of
  SIGIR '20}}. \bibinfo{pages}{329--338}.
\newblock


\bibitem[\protect\citeauthoryear{Beel, Genzmehr, Langer, N\"{u}rnberger, and
  Gipp}{Beel et~al\mbox{.}}{2013}]%
        {Beel:2013:CAO}
\bibfield{author}{\bibinfo{person}{Joeran Beel}, \bibinfo{person}{Marcel
  Genzmehr}, \bibinfo{person}{Stefan Langer}, \bibinfo{person}{Andreas
  N\"{u}rnberger}, {and} \bibinfo{person}{Bela Gipp}.}
  \bibinfo{year}{2013}\natexlab{}.
\newblock \showarticletitle{{A Comparative Analysis of Offline and Online
  Evaluations and Discussion of Research Paper Recommender System Evaluation}}.
  In \bibinfo{booktitle}{\emph{Proc. of RepSys '13 workshop}}.
  \bibinfo{pages}{7--14}.
\newblock


\bibitem[\protect\citeauthoryear{Garcin, Faltings, Donatsch, Alazzawi, Bruttin,
  and Huber}{Garcin et~al\mbox{.}}{2014}]%
        {Garcin:2014:OOE}
\bibfield{author}{\bibinfo{person}{Florent Garcin}, \bibinfo{person}{Boi
  Faltings}, \bibinfo{person}{Olivier Donatsch}, \bibinfo{person}{Ayar
  Alazzawi}, \bibinfo{person}{Christophe Bruttin}, {and} \bibinfo{person}{Amr
  Huber}.} \bibinfo{year}{2014}\natexlab{}.
\newblock \showarticletitle{{Offline and Online Evaluation of News Recommender
  Systems at Swissinfo.Ch}}. In \bibinfo{booktitle}{\emph{Proc. of RecSys
  '14}}. \bibinfo{pages}{169--176}.
\newblock


\bibitem[\protect\citeauthoryear{Hofmann, Li, and Radlinski}{Hofmann
  et~al\mbox{.}}{2016}]%
        {Hofmann:2016:OEI}
\bibfield{author}{\bibinfo{person}{Katja Hofmann}, \bibinfo{person}{Lihong Li},
  {and} \bibinfo{person}{Filip Radlinski}.} \bibinfo{year}{2016}\natexlab{}.
\newblock \showarticletitle{Online Evaluation for Information Retrieval}.
\newblock \bibinfo{journal}{\emph{Found. Trends Inf. Retr.}}
  \bibinfo{volume}{10}, \bibinfo{number}{1} (\bibinfo{date}{June}
  \bibinfo{year}{2016}), \bibinfo{pages}{1--117}.
\newblock


\bibitem[\protect\citeauthoryear{Hopfgartner, Balog, Lommatzsch, Kelly, Kille,
  Schuth, and Larson}{Hopfgartner et~al\mbox{.}}{2019}]%
        {Hopfgartner:2019:CEL}
\bibfield{author}{\bibinfo{person}{Frank Hopfgartner},
  \bibinfo{person}{Krisztian Balog}, \bibinfo{person}{Andreas Lommatzsch},
  \bibinfo{person}{Liadh Kelly}, \bibinfo{person}{Benjamin Kille},
  \bibinfo{person}{Anne Schuth}, {and} \bibinfo{person}{Martha Larson}.}
  \bibinfo{year}{2019}\natexlab{}.
\newblock \showarticletitle{{Continuous Evaluation of Large-Scale Information
  Access Systems: A Case for Living Labs}}.
\newblock In \bibinfo{booktitle}{\emph{Information Retrieval Evaluation in a
  Changing World - Lessons Learned from 20 Years of {CLEF}}}.
  \bibinfo{series}{The Information Retrieval Series},
  Vol.~\bibinfo{volume}{41}. \bibinfo{publisher}{Springer},
  \bibinfo{pages}{511--543}.
\newblock


\bibitem[\protect\citeauthoryear{Hopfgartner, Brodt, Seiler, Kille, Lommatzsch,
  Larson, Turrin, and Ser{\'{e}}ny}{Hopfgartner et~al\mbox{.}}{2015}]%
        {HopfgartnerBSKL15}
\bibfield{author}{\bibinfo{person}{Frank Hopfgartner}, \bibinfo{person}{Torben
  Brodt}, \bibinfo{person}{Jonas Seiler}, \bibinfo{person}{Benjamin Kille},
  \bibinfo{person}{Andreas Lommatzsch}, \bibinfo{person}{Martha Larson},
  \bibinfo{person}{Roberto Turrin}, {and} \bibinfo{person}{Andr{\'{a}}s
  Ser{\'{e}}ny}.} \bibinfo{year}{2015}\natexlab{}.
\newblock \showarticletitle{{Benchmarking News Recommendations: The CLEF
  NewsREEL Use Case}}.
\newblock \bibinfo{journal}{\emph{{SIGIR} Forum}} \bibinfo{volume}{49},
  \bibinfo{number}{2} (\bibinfo{year}{2015}), \bibinfo{pages}{129--136}.
\newblock


\bibitem[\protect\citeauthoryear{Jagerman, Balog, and Rijke}{Jagerman
  et~al\mbox{.}}{2018}]%
        {Jagerman:2018:OLL}
\bibfield{author}{\bibinfo{person}{Rolf Jagerman}, \bibinfo{person}{Krisztian
  Balog}, {and} \bibinfo{person}{Maarten~De Rijke}.}
  \bibinfo{year}{2018}\natexlab{}.
\newblock \showarticletitle{OpenSearch: Lessons Learned from an Online
  Evaluation Campaign}.
\newblock \bibinfo{journal}{\emph{J. Data and Information Quality}}
  \bibinfo{volume}{10}, \bibinfo{number}{3}, Article \bibinfo{articleno}{13}
  (\bibinfo{date}{Sept.} \bibinfo{year}{2018}),
  \bibinfo{numpages}{13:1--13:15}~pages.
\newblock


\bibitem[\protect\citeauthoryear{Monroe}{Monroe}{2018}]%
        {Monroe:2018:AEY}
\bibfield{author}{\bibinfo{person}{Don Monroe}.}
  \bibinfo{year}{2018}\natexlab{}.
\newblock \showarticletitle{{AI, Explain Yourself}}.
\newblock \bibinfo{journal}{\emph{Commun. ACM}} \bibinfo{volume}{61},
  \bibinfo{number}{11} (\bibinfo{date}{oct} \bibinfo{year}{2018}),
  \bibinfo{pages}{11--13}.
\newblock


\bibitem[\protect\citeauthoryear{Schuth}{Schuth}{2016}]%
        {Schuth:2016:SEL}
\bibfield{author}{\bibinfo{person}{Anne Schuth}.}
  \bibinfo{year}{2016}\natexlab{}.
\newblock \emph{\bibinfo{title}{Search Engines that Learn from Their Users}}.
\newblock \bibinfo{thesistype}{Ph.D. Dissertation}. \bibinfo{school}{University
  of Amsterdam}.
\newblock


\bibitem[\protect\citeauthoryear{Schuth, Balog, and Kelly}{Schuth
  et~al\mbox{.}}{2015}]%
        {Schuth:2015:OLL}
\bibfield{author}{\bibinfo{person}{Anne Schuth}, \bibinfo{person}{Krisztian
  Balog}, {and} \bibinfo{person}{Liadh Kelly}.}
  \bibinfo{year}{2015}\natexlab{}.
\newblock \showarticletitle{{Overview of the Living Labs for Information
  Retrieval Evaluation (LL4IR) CLEF Lab 2015}}. In
  \bibinfo{booktitle}{\emph{Proc. of CLEF'15}}. \bibinfo{pages}{484--496}.
\newblock


\bibitem[\protect\citeauthoryear{Wu, Williams, Chen, Khabsa, Caragea, Ororbia,
  Jordan, and Giles}{Wu et~al\mbox{.}}{2014}]%
        {wu-citeseerx-2014}
\bibfield{author}{\bibinfo{person}{Jian Wu}, \bibinfo{person}{Kyle Williams},
  \bibinfo{person}{Hung-Hsuan Chen}, \bibinfo{person}{Madian Khabsa},
  \bibinfo{person}{Cornelia Caragea}, \bibinfo{person}{Alexander Ororbia},
  \bibinfo{person}{Douglas Jordan}, {and} \bibinfo{person}{C.~Lee Giles}.}
  \bibinfo{year}{2014}\natexlab{}.
\newblock \showarticletitle{CiteSeerX: AI in a Digital Library Search Engine}.
  In \bibinfo{booktitle}{\emph{Proc. of AAAI '14}}.
  \bibinfo{pages}{2930--2937}.
\newblock


\bibitem[\protect\citeauthoryear{Zhang and Chen}{Zhang and Chen}{2020}]%
        {Zhang:2020:ERS}
\bibfield{author}{\bibinfo{person}{Yongfeng Zhang} {and} \bibinfo{person}{Xu
  Chen}.} \bibinfo{year}{2020}\natexlab{}.
\newblock \showarticletitle{Explainable Recommendation: {A} Survey and New
  Perspectives}.
\newblock \bibinfo{journal}{\emph{Found. Trends Inf. Retr.}}
  \bibinfo{volume}{14}, \bibinfo{number}{1} (\bibinfo{year}{2020}),
  \bibinfo{pages}{1--101}.
\newblock


\end{thebibliography}

\end{document}